\documentclass[twocolumn,showpacs,amsfonts,aps,prl,floatfix,%
superscriptaddress]{revtex4} 
\usepackage{amsmath}
\usepackage{bm}
\usepackage{graphicx}

\newcommand{\be}[1]{\begin{equation}\label{#1}}
\newcommand{\ee}{\end{equation}}

\newcommand{\eq}{{\,=\,}}

\usepackage{color}

\begin{document}

%%%%%%%%%%%%%%%%%%%%%%%%Front Matter%%%%%%%%%%%%%%%%%%%%%%%%%%%%%%%%%%
%%%%%%%%%%%%%%%%%%%%%%%%%%%%%%%%%%%%%%%%%%%%%%%%%%%%%%%%%%%%%%%%%%%%%%

\title{Hadronic dissipative effects on elliptic flow in 
ultrarelativistic heavy-ion collisions}
\date{\today}

\author{Tetsufumi Hirano}
\email[Correspond to\ ]{hirano@phys.columbia.edu}
\affiliation{Department of Physics, Columbia University, 538 West 
120th Street, New York, NY 10027, USA}
\author{Ulrich Heinz}
\affiliation{Department of Physics, Ohio State University, 191 West
  Woodruff Avenue, Columbus, OH 43210, USA}
\author{Dmitri Kharzeev}
\affiliation{Nuclear Theory Group, Physics Department, Brookhaven 
             National Laboratory, Upton, NY 11973-5000, USA}
\author{Roy Lacey}
\affiliation{Department of Chemistry, SUNY Stony Brook, Stony Brook, 
NY 11794-3400, USA}
\author{Yasushi Nara}
\affiliation{Institut f\"ur Theoretische Physik, 
J.\,W.\,Goethe-Universit\"at, Max\,v.\,Laue\,Str.\,1, D-60438 Frankfurt,
Germany}

\begin{abstract}
We study the elliptic flow coefficient $v_2(\eta,b)$ in Au+Au collisions 
at $\sqrt{s}=200\,A$\,GeV as a function of pseudorapidity $\eta$ and 
impact parameter $b$. Using a hybrid approach which combines early ideal 
fluid dynamical evolution with late hadronic rescattering, we demonstrate 
strong dissipative effects from the hadronic rescattering stage on the 
elliptic flow. With Glauber model initial conditions, hadronic
dissipation is shown to be sufficient to fully explain the differences
between measured $v_2$ values and ideal hydrodynamic predictions. 
Initial conditions based on the Color Glass Condensate model generate
larger elliptic flow and seem to require additional dissipation during 
the early quark-gluon plasma stage in order to achieve agreement with 
experiment.
\end{abstract}


\maketitle

One of the important new discoveries made at the Relativistic Heavy Ion 
Collider (RHIC) is the large elliptic flow $v_2$ in non-central Au+Au 
collisions \cite{experiments}. At the highest RHIC energy of 
$\sqrt{s}=200\,A$\,GeV, the observed $v_2$ values near midrapidity 
($|\eta|{\,\alt\,}1$), for not too large impact parameters 
($b{\,\alt\,}7$\,fm) and transverse momenta ($p_T{\,\alt\,}1.5$\,GeV/$c$), 
agree with predictions from ideal fluid dynamics \cite{QGP3}, including 
\cite{STARv2,PHENIXv2} the predicted dependence of $v_2$ on the 
transverse momentum $p_T$ and hadron rest masses \cite{HKHRV01}. From 
these observations it has been concluded \cite{HK02} that in these 
collisions a quark-gluon plasma (QGP) is created which thermalizes on 
a very rapid time scale $\tau_{\rm therm}< 1$\,fm/$c$ and subsequently 
evolves as an almost ideal fluid with exceptionally low viscosity.

On the other hand, the ideal fluid dynamical description gradually
breaks down as one studies collisions at larger impact parameters 
and at lower energies \cite{NA49v2sys} or moves away from midrapidity 
\cite{PHOBOSv2eta,Hiranov2eta,HT02,fn1}. This has been attributed 
alternatively to incomplete thermalization of the QGP during the 
early stages of the expansion \cite{HK04} (``early viscosity'') 
and/or to dissipative effects during the late hadronic expansion 
stage \cite{Teaney,Gyulassy_Kemer,HG05} (``late viscosity'').
It has recently been argued \cite{Son_visc,Gyulassy_Kemer} that 
quantum mechanics imposes a lower limit on the shear viscosity 
{\em of any medium}, but that the shear viscosity of the QGP can 
not exceed this lower limit by a large factor \cite{Teaney_visc}. 
On the other hand, qualitative arguments were presented in 
Ref.~\cite{HG05} which emphasize the importance of hadronic 
dissipation and support a picture of a ``nearly perfect fluid 
strongly coupled QGP (sQGP) core and highly dissipative hadronic 
corona" in ultrarelativistic heavy-ion collisions. 
The importance
of viscous effects for a successful description of RHIC data on 
$v_2$ and $v_4$ was also emphasized in \cite{Bhalerao}, although 
this work left it open whether the corresponding lack of 
thermalization occurs mostly at the beginning or towards the end of 
the expansion phase. 
In the present paper we explore this issue more 
quantitatively, by trying to answer the question {\em how much} of the 
observed deviation of $v_2$ from the ideal fluid prediction can be 
attributed to ``late viscosity'' in the dissipative hadronic phase, 
and whether or not significant additional dissipative effects during 
the early QGP stage are required for a quantitative understanding of 
the data. 

Our study is based on a comparison of a hybrid model, combining an 
ideal fluid dynamical QGP stage with a realistic kinetic description 
of the hadronic stage (hadron cascade), with data on the centrality 
and rapidity dependence of the $p_T$-integrated elliptic flow 
$v_2(\eta,b)$ for charged hadrons \cite{data}. We find that with 
Glauber model initial conditions \cite{Kolb:2001qz}, suitably 
generalized to account for the longitudinal structure of the initial 
fireball \cite{AG05}, hadronic dissipation is sufficient to explain 
the data. On the other hand, initial conditions based on the Color 
Glass Condensate (CGC) model \cite{KLN01,HN04} lead to larger elliptic 
flows which overpredict the data unless one additionally assumes that 
the early QGP stage possesses significant shear viscosity, too, or 
that the QGP equation of state is significantly softer than usually 
assumed. Our analysis points to a need for a better understanding of 
the initial conditions in heavy-ion collisions if one hopes to use 
experimental data to constrain the QGP viscosity and equation of state.

A (1+1)-dimensional hydro+cascade model was first proposed in 
Ref.~\cite{BassDumitru}, putting emphasis on radial flow in heavy-ion
collisions. It was later extended to 2+1 dimensions for the study of 
elliptic flow near midrapidity~\cite{Teaney,TLS01}. By combining a 
hydrodynamic description of the early expansion stage with a hadron 
transport model at the end we can implement a realistic treatment 
of the freeze-out process and of viscous effects during the final 
hadronic phase. Here we extend the above models to full 
(3+1)-dimensional hydrodynamics \cite{NonakaBass}, in order to be 
able to study the rapidity dependence of elliptic flow. Let us briefly 
summarize our model. For the hydrodynamic stage, we solve the 
conservation laws $\partial_\mu T^{\mu\nu}\eq0$ with the ideal fluid 
decomposition $T^{\mu\nu}\eq(e{+}p)u^\mu u^\nu - pg^{\mu\nu}$ (where 
$e$ and $p$ are energy density and pressure and $u^\mu$ is the fluid 
4-velocity) in Bjorken coordinates $(\tau,\bm{x}_{\perp},\eta_s)$ 
\cite{Hiranov2eta}. We neglect the finite (but at RHIC energy very 
small) net baryon density. A massless ideal parton gas equation of 
state (EOS) is employed in the QGP phase ($T{\,>\,}T_c\eq170$ MeV) 
while a hadronic resonance gas model is used at $T{\,<\,}T_c$. When 
we use the hydrodynamic code all the way to final decoupling, we 
take into account \cite{HT02} chemical freezeout of the hadron 
abundances at $T_{\mathrm{ch}}\eq170$ MeV, separated from thermal 
freezeout of the momentum spectra at a lower decoupling temperature 
$T_{\mathrm{dec}}$, as required to reproduce the experimentally 
measured yields \cite{BMRS01}. 

For the hydro+cascade description, a hadronic transport model 
JAM~\cite{jam} is employed for the late stage of the expansion.
JAM simulates nuclear collisions by individual hadron-hadron 
collisions. Soft hadron production in hadron-hadron scattering
is modeled by exciting hadronic resonances and color strings.
Color strings decay into hadrons after their formation time 
($\tau{\,\sim\,}1$\,fm/$c$) according to the Lund string model 
PYTHIA~\cite{Sjostrand}. Leading hadrons which contain original 
constituent quarks can scatter within their formation time 
with other hadrons assuming additive quark cross sections
\cite{RQMDUrQMD}. In the current study, it is initialized
with output from the above (3+1)-dimensional hydrodynamics by 
using the Cooper-Frye formalism \cite{CF} (rejecting 
backward going particles) \cite{Teaney,TLS01}. We switch 
from hydrodynamics to the cascade approach at the switching 
temperature $T_{\mathrm{sw}}\eq169$ MeV, i.e. just below the 
hadronization phase transition.

We here study two types of initial conditions for the evolution. The first,
which we call ``modified BGK initial condition'' \cite{BGK,AG05}, assumes
an initial entropy distribution of massless partons 
according to
\begin{eqnarray}
\label{eq:1}
 \frac{dS}{d\eta_s d^2x_\perp} &=& 
 {C\over 1+\alpha}\, \theta\bigl(Y_b{-}|\eta_s|\bigr)\, f^{pp}(\eta_s)
\nonumber\\
 &\times& \Biggl[\alpha 
  \left(\frac{Y_b{-}\eta_s}{Y_b}\,\frac{dN^A_{\rm part}}{d^2x_\perp} 
      + \frac{Y_b{+}\eta_s}{Y_b}\,\frac{dN^B_{\rm part}}{d^2x_\perp}\right)
\nonumber\\
 && \  + (1{-}\alpha)\, \frac{dN_{\rm coll}}{d^2x_\perp}\Biggr],
\end{eqnarray}
where $\eta_s\eq\frac{1}{2}\ln[(t{+}z)/(t{-}z)]$ is the space-time
rapidity, $\bm{x}_\perp\eq(x,y)$ is the position transverse to the 
beam axis, $C\eq24.0$ is chosen to reproduce the measured charged hadron 
multiplicity in central collisions at midrapidity \cite{PHOBOS_Nch},
$Y_b$ is the beam rapidity, and $f^{pp}$ is a suitable parametrization 
of the shape of rapidity distribution in $pp$ collisions,
\begin{equation}\label{eq:2}
  f^{pp}(\eta_s) = \exp\left[-\theta(|\eta_s|{-}\Delta\eta)\,
  \frac{(|\eta_s|{-}\Delta\eta)^2}{\sigma_\eta^2}\right],
\end{equation}
with $\Delta\eta\eq1.3$ and $\sigma_\eta\eq2.1$,
which are so chosen as to reproduce the measured pseudorapidity
distributions for charged hadrons \cite{PHOBOS_dNdeta}. 
$N^{A,B}_{\rm part}$ and $N_{\rm coll}$
are the number of wounded nucleons in the two nuclei and the number
of binary nucleon-nucleon collisions, respectively, as calculated
from the Glauber model nuclear thickness function $T_{A,B}(\bm{x}_\perp)$
\cite{Kolb:2001qz},
\begin{eqnarray}
\label{eq:3}
\frac{dN^A_{\rm part}}{d^2x_\perp} &=& 
T_A(r_+)\left[1 - 
\left(1-\frac{\sigma_{NN}^{\rm in}\,T_B(r_-)}{B}\right)^B
        \right],
\\  
\label{eq:4}
\frac{dN^B_{\rm part}}{d^2x_\perp} &=& 
T_B(r_-)\left[1 - \left(1-\frac{\sigma_{NN}^{\rm in}\,T_A(r_+)}{A}\right)^A
        \right],
\\  
\label{eq:5}
\frac{dN_{\rm coll}}{d^2x_\perp} &=& 
\sigma_{NN}^{\rm in}\,T_A(r_+)\,T_B(r_-),
\end{eqnarray}
with the inelastic nucleon-nucleon cross section 
$\sigma_{NN}^{\rm in} = 42$\,mb 
and $r_\pm\eq\bigl[\left(x{\pm}\frac{1}{2}b\right)^2+y^2\bigr]^{1/2}$ 
(where $b$ is the impact parameter). The soft/hard
fraction $\alpha\eq0.85$ was adjusted to reproduce the measured 
centrality dependence \cite{PHOBOS_Nch} of the charged hadron multiplicity 
at midrapidity. At $\eta_s\eq0$, Eq.~(\ref{eq:1}) reduces to 
$dS/(d\eta_s d^2x_\perp) \propto [\alpha (N^A_{\rm part}{+}N^B_{\rm part}) 
+ (1{-}\alpha)N_{\rm coll}]/(1{+}\alpha)$ \cite{KH05}; 
this parameterization is equivalent to the one used in Ref.~\cite{KLN01}, 
$ \sim \frac{1{-}x}{2} (N^A_{\rm part}{+}N^B_{\rm part}) + x N_{\rm coll}$, 
with $x\eq\frac{1{-}\alpha}{1{+}\alpha}$. 
From Eq.~(\ref{eq:1}), we can 
compute the entropy density at the initial time $\tau_0\eq0.6$\,fm/$c$ 
\cite{QGP3} of the hydrodynamic evolution, 
$s(\tau_0,\bm{x}_\perp, \eta_s)\eq{dS}/(\tau_0 d\eta_s d^2x_\perp)$,
which provides the initial energy density and pressure distributions 
through the tabulated EOS described above.

The second type of initial conditions is based on the CGC model
\cite{MV}. Specifically, we use the Kharzeev-Levin-Nardi (KLN) 
approach \cite{KLN01} in the version previously employed in \cite{HN04}. 
In this approach, the energy distribution of produced gluons with 
rapidity $y$ is given by the $k_T$-factorization formula \cite{GLR83}
\begin{eqnarray}
   \frac{dE_T}{d^2x_{\perp}dy}&=&
   \frac{4\pi^2N_c}{N_c^2-1} \int\frac{d^2p_T}{p_T}
   \int^{p_T} \frac{d^2k_T}{4} \alpha_s(Q^2)   \nonumber\\
  &\times&      \phi_A(x_1,(\bm{p}_T{+}\bm{k}_T)^2/4;\bm{x}_\perp) \nonumber\\
  &\times&      \phi_B(x_2,(\bm{p}_T{-}\bm{k}_T)^2/4;\bm{x}_\perp), 
\label{eq:ktfac}
\end{eqnarray}
where $x_{1,2}{\eq}p_T\exp(\pm y)/\sqrt{s}$ and $p_T$ is the transverse 
momentum of the produced gluons. We choose an upper limit of 3\,GeV/$c$
for the $p_T$ integration. For the unintegrated gluon distribution 
function we use
\begin{equation}
\label{eq:uninteg}
  \phi_A(x,k^2_T;\bm{x}_\perp)
  =\left\{\begin{array}{l}
   \frac{\kappa C_F}{2\pi^3\alpha_s(Q^2_s)}\frac{Q_s^2}{Q_s^2+\Lambda^2}, 
   \,                      \quad  k_T\,\leq\,Q_s, \\
   \frac{\kappa C_F}{2\pi^3\alpha_s(Q^2_s)}\, \frac{Q^2_s}{k^2_T+\Lambda^2},
              \quad k_T\,>\,Q_s,
\end{array}
\right.
\end{equation}
where $C_F\eq\frac{N_c^2{-}1}{2N_c}$ and $Q_s$ denotes the saturation 
momentum. We introduce a small regulator $\Lambda=0.2$\,GeV/$c$ in 
order to have a smooth distribution in the forward rapidity region
$|y|{\,>\,}4.5$ at RHIC (other regions are not affected by introducing 
this small regulator). The overall normalization $\kappa$ is 
determined by fitting the multiplicity of charged hadron at 
midrapidity at $\sqrt{s_{NN}}=200$ GeV for the most central collisions.
The saturation momentum $Q_s$ of nucleus $A$ in $A{+}B$ collisions, 
needed in the function $\phi_A$, is obtained by solving the following 
implicit equation at fixed $x$ and $\bm{x}_{\perp}$:
\begin{equation}
\label{eq:saturation}
   Q^2_s(x, \bm{x}_{\perp}) = \frac{2\pi^2}{C_F}
   \,\alpha_s(Q^2_s)\,x G(x,Q^2_s)\,
   \frac{dN^A_{\mathrm{part}}}{d^2 x_\perp}.
\end{equation}
An analogous equation holds for the saturation momentum of nucleus $B$
in $\phi_B$. For the gluon distribution function $G$ inside a nucleon 
we take the 
simple ansatz \cite{KLN01}
\begin{equation}
\label{eq:xG}
  xG(x,Q^2) = K\ln\left( \frac{Q_s^2 + \Lambda^2}
               {\Lambda_{\mathrm{QCD}}^2}\right)
	       x^{-\lambda} (1-x)^4
\end{equation}
with $\Lambda\eq\Lambda_{\mathrm{QCD}}\eq0.2$\,GeV. We choose 
$K\eq0.7$ and $\lambda\eq0.2$ so that the average saturation 
momentum in the transverse plane yields 
$\langle Q_s^2(x{=}0.01)\rangle{\,\sim\,}2.0$\,GeV$^2/c^2$
in central 200\,$A$\,GeV Au+Au collisions at RHIC. For the running 
coupling constant $\alpha_s$ in Eq.~(\ref{eq:saturation}) we use
the standard perturbative one--loop formula, 
but introducing a cut-off in the infra-red region of
small $Q_s$ (i.e. near the surface of the nuclear
overlap region where the produced gluon density is low) by limiting 
the coupling constant to $\alpha_s{\,\leq\,}0.5$.
We can obtain the energy density distribution at time $\tau_0$ 
from Eq.~(\ref{eq:ktfac}), 
$e(\tau_0,\bm{x}_\perp,\eta_s){\eq}dE_T/(\tau_0 d\eta_s d^2x_\perp)$,
where $y$ is identified with $\eta_s$, and use this as the initial 
distribution for the hydrodynamic evolution.

%
%%%%%%%%%%%%%%%%%%%%%%%%%%% Fig. 1 %%%%%%%%%%%%%%%%%%%%%%%%%%%%%%%%%%%%%%%
\begin{figure}[t]
\includegraphics[width = \linewidth,clip]{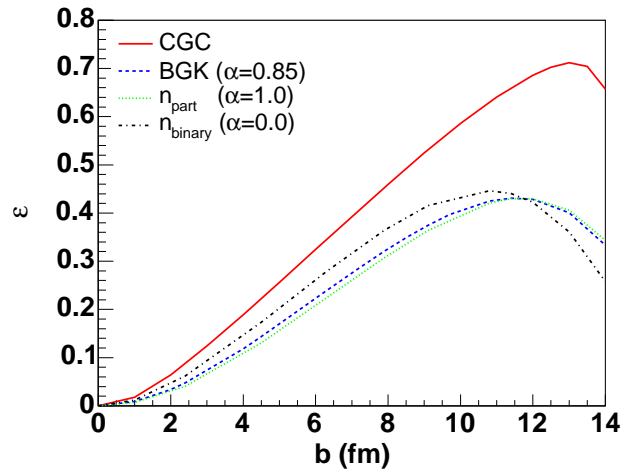}
\caption{(Color online) Initial spatial eccentricity $\varepsilon\eq\frac
{\langle y^2{-}x^2\rangle}{\langle y^2{+}x^2\rangle}$ at midrapidity
as a function of impact parameter $b$, for 200\,$A$\,GeV Au+Au collisions 
with CGC (solid red) and BGK (dashed blue) initial conditions. For 
comparison we also show initial 
conditions where the initial parton density at midrapidity scales 
with the transverse density of wounded nucleons (dotted green)
and of binary collisions (dash-dotted black) \cite{Kolb:2001qz}.
%vspace*{-8mm}
}
\label{F1}
\end{figure}
%%%%%%%%%%%%%%%%%%%%%%%%%%%%%%%%%%%%%%%%%%%%%%%%%%%%%%%%%%%%%%%%%%%%%%%%%%%
%

In Fig.~\ref{F1} we show the initial eccentricity $\varepsilon\eq\frac
{\langle y^2{-}x^2\rangle}{\langle y^2{+}x^2\rangle}$ of the source at
midrapidity ($\eta_s\eq0$) for our two models for the initial conditions.
Here $\langle \cdots \rangle$ represents the average taken with respect 
to the initial energy density distribution 
$e(\tau_0,{\bm x}_\perp, \eta_s\eq0)$.
While the BGK model interpolates between the binary collision and wounded 
nucleon scaling curves (being closer to the latter), CGC initial conditions
are seen to give much larger eccentricities. This can be traced back to
a steeper drop of the energy density profile near the edge in the CGC 
model. In ideal fluid dynamics, the larger eccentricities $\varepsilon$ 
result in larger elliptic flow coefficients $v_2$ \cite{Ollitrault}. 

%
%%%%%%%%%%%%%%%%%%%%%%%%%%% Fig. 2 %%%%%%%%%%%%%%%%%%%%%%%%%%%%%%%%%%%%%%%
\begin{figure}[t]
\includegraphics[width = \linewidth,clip]{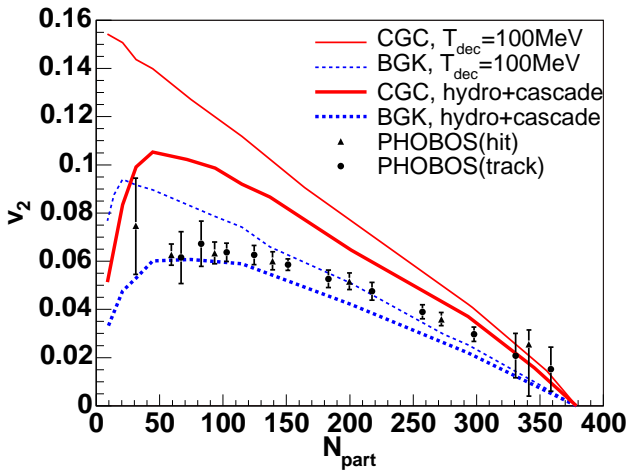}
\caption{(Color online) $p_T$-integrated elliptic flow for charged hadrons 
at midrapidity ($\mid\eta\mid < 1$) 
from 200\,$A$\,GeV Au+Au collisions, as a function of 
the number $N_{\rm part}$ of participating nucleons. The thin lines 
show the prediction from ideal fluid dynamics with a freeze-out 
temperature $T_{\rm dec}\eq100$\,MeV, for CGC (solid red) and BGK 
(dashed blue) initial conditions. The thick lines (solid red for CGC and
dashed blue for BGK initial conditions) show the corresponding results 
from the hydro+cascade hybrid model. The data are from the PHOBOS 
Collaboration \cite{data}.
%vspace*{-8mm}
}
\label{F2}
\end{figure}
%%%%%%%%%%%%%%%%%%%%%%%%%%%%%%%%%%%%%%%%%%%%%%%%%%%%%%%%%%%%%%%%%%%%%%%%%%%
%

This is shown by the thin lines in Fig.~\ref{F2} which compare ideal
fluid dynamical calculations with RHIC data from the PHOBOS 
collaboration \cite{data}. Note that the hydrodynamic calculations 
shown here use an EOS in which the hadron abundances are held fixed 
below $T_c$ at their chemical freeze-out values established during 
the hadronization process \cite{HT02}. For this EOS it is known from
earlier hydrodynamic studies \cite{HT02,KR03} that the slope of pion 
$p_T$ spectrum stalls during the hadronic evolution. Without an 
additional transverse velocity kick already at the beginning of
the hydrodynamic stage \cite{KR03} it reproduces the pion data from 
central collisions at midrapidity only up to $\sim$1\,GeV/$c$, falling 
off too steeply at larger $p_T$. In contrast, the proton $p_T$ spectrum 
continues to flatten during the hadronic stage, due to continued 
build-up of radial flow, reproducing the proton data up to 
$\sim$2\,GeV/$c$ if one chooses $T_{\rm dec}\eq100$\,MeV as
the freeze-out temperature. The hybrid hydro+cascade model gives 
almost the same proton $p_T$ spectrum as ideal fluid dynamics with 
$T_{\rm dec}\eq100$\,MeV, but further flattens the pion spectrum at 
larger $p_T$, thereby improving the agreement with the pion data 
up to $\sim$2\,GeV/$c$. This can be understood as a consequence of
shear viscosity in the hadronic phase, with the fast longitudinal
expansion generating positive additional viscous pressure components
in the transverse directions whose effects on the $p_T$ spectrum 
grow quadratically with $p_T$ \cite{Teaney_visc} and decrease with 
increasing hadron mass. The agreement with the measured $p_T$ spectra
is slightly better for CGC than for BGK initial conditions.    

%
%%%%%%%%%%%%%%%%%%%%%%%%%%% Fig. 3 %%%%%%%%%%%%%%%%%%%%%%%%%%%%%%%%%%%%%%
\begin{figure}[t]
\includegraphics[width = \linewidth,clip]{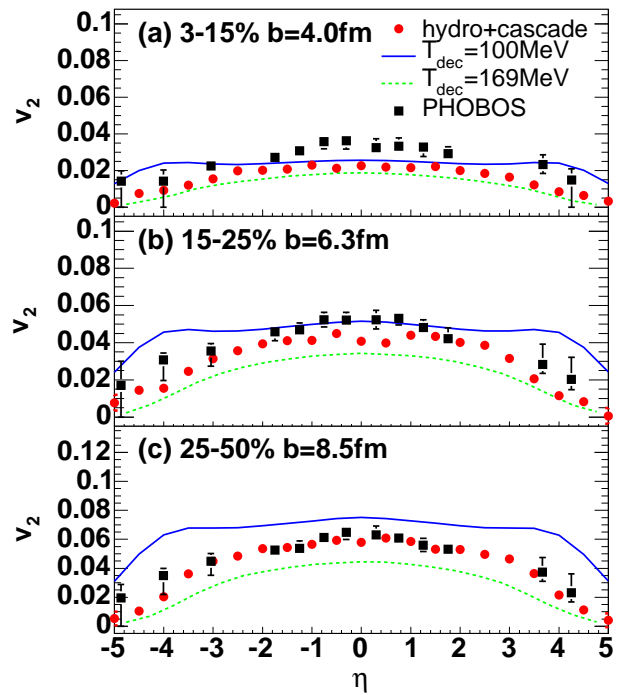}
\caption{(Color online)
The pseudorapidity dependence of $v_2$ for charged hadrons in 
(a) central (3-15\%),
(b) semicentral (15-25\%), and 
(c) peripheral (25-50\%)
Au+Au collisions at $\sqrt{s}\eq200\,A$\,GeV. The corresponding 
impact parameters are, respectively, $b=4.0, 6.3,$ and 8.5\,fm.
The hydrodynamic evolution is initialized with modified BGK 
initial conditions. The lines show the predictions from ideal
fluid dynamics with $T_\mathrm{dec}\eq100$ MeV (solid blue) and 
$T_\mathrm{dec}\eq169$ MeV (dashed green). The red circles show 
the corresponding results from the hydro+cascade hybrid model.
The black squares are measurements by the PHOBOS Collaboration 
\cite{data}.
%vspace*{-8mm}
}
\label{F3}
\end{figure}
%%%%%%%%%%%%%%%%%%%%%%%%%%%%%%%%%%%%%%%%%%%%%%%%%%%%%%%%%%%%%%%%%%%%%%%%%%%
%
%
%%%%%%%%%%%%%%%%%%%%%%%%%%% Fig. 4 %%%%%%%%%%%%%%%%%%%%%%%%%%%%%%%%%%%%%%
\begin{figure}[t]
\includegraphics[width = \linewidth,clip]{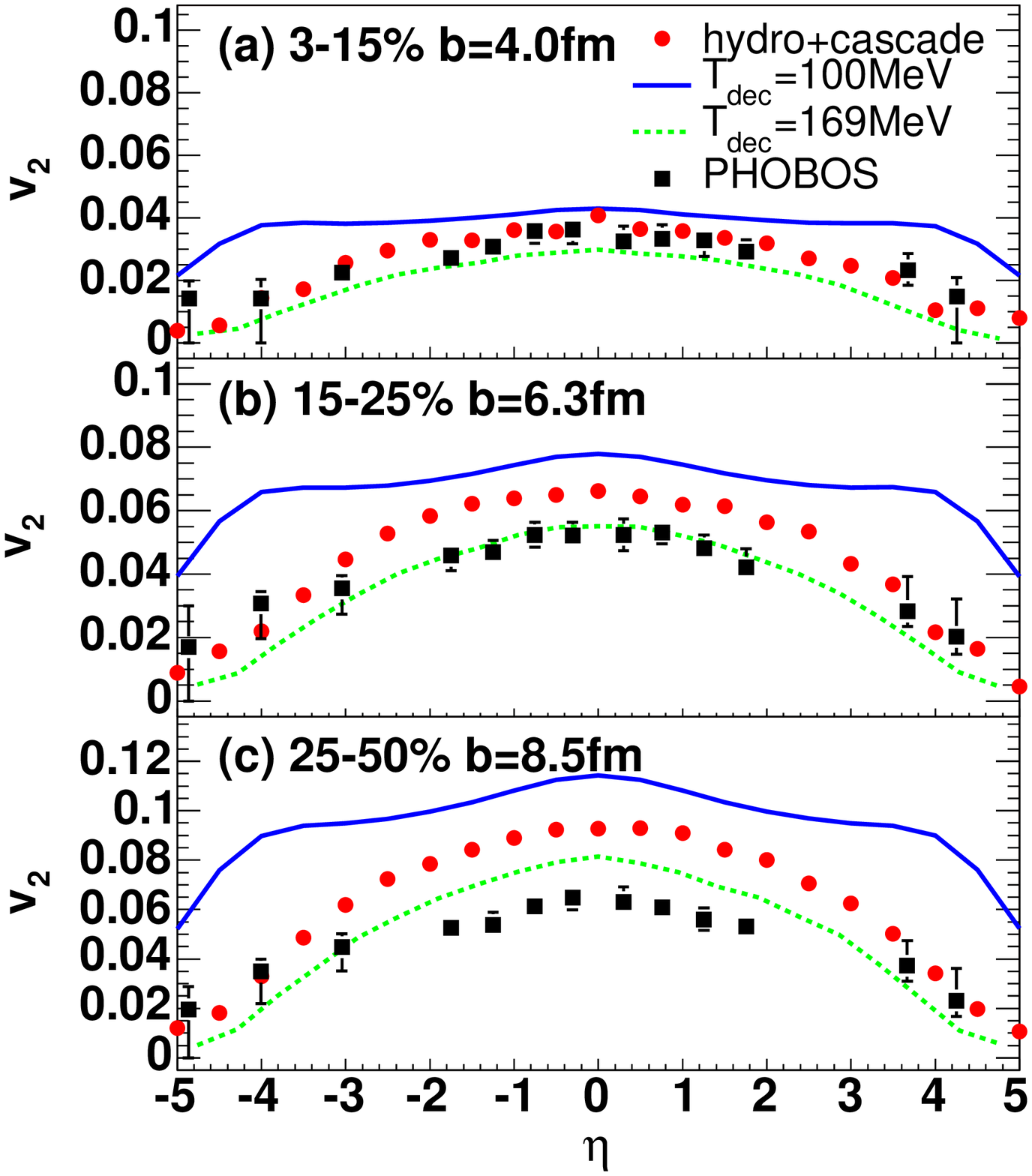}
\caption{(Color online) Same as Fig.~\ref{F3}, except for using 
CGC instead of BGK initial conditions.
%vspace*{-8mm}
}
\label{F4}
\end{figure}
%%%%%%%%%%%%%%%%%%%%%%%%%%%%%%%%%%%%%%%%%%%%%%%%%%%%%%%%%%%%%%%%%%%%%%%%%%%
%

As explained in \cite{HG05}, the chemically frozen hadronic EOS
results in smaller $v_2$ values than for a hadronic EOS which assumes
hadronic chemical equilibrium all the way down to kinetic freeze-out 
at $T_{\rm dec}$. Correspondingly, our curve for BGK initial conditions
lies below the one shown in Fig.~2 of \cite{data} which uses the latter
EOS. It is noteworthy that for central and semicentral collisions
the data seem to lie consistently somewhat above the hydrodynamic 
predictions with BGK initial conditions. While at first sight this 
seems to argue against the validity of the BGK model, it must be
noted that event-by-event fluctuations in the geometry of the nuclear
overlap region, which we don't take into account \cite{fn2}, tend to 
significantly increase the {\em measured} $v_2$ in collisions with 
small impact parameters (large $N_{\rm part}$ values) \cite{MS03}.

The difference between the eccentricities given by the BGK and CGC 
initial conditions may seem surprising in view of the fact that the 
centrality dependence of hadron multiplicities in both approaches 
can be made numerically very similar by a proper choice of parameter 
$\alpha$ in Eq.~(\ref{eq:1}) \cite{KLN01}. When parameterized in terms 
of Eq.~(\ref{eq:1}), the main prediction of the CGC approach is the 
near independence of $\alpha$ on the collision energy, which is 
confirmed by the data. The reason for the big difference in the 
eccentricities stems from the different entropy profiles predicted 
by the two approaches, especially in the regions where the density 
of produced particles is relatively small. While these differences 
contribute little to the total observed multiplicity, they appear 
quite important for the evaluation of eccentricity. The application 
of the CGC approach in the region of small parton density is of 
course questionable, and a better theoretical understanding of the 
transition from high to low density regimes is clearly needed.   

The thick lines show the centrality dependence of $v_2$ from the 
hydro+cascade hybrid model. Whereas with BGK initial conditions
the dissipative effects of the hadronic phase in the cascade model
reduce the purely hydrodynamic $v_2$ sufficiently to bring the 
theory in agreement with the data even for peripheral collisions, 
CGC initial conditions, with their larger eccentricities, cause so 
much elliptic flow that even the hybrid model overpredicts the data
significantly. This is true for both the $p_T$-integrated elliptic
flow (shown in Fig.~\ref{F2}) and its $p_T$-slope, $dv_2(p_T)/dp_T$, 
at midrapidity, for both pions and protons (to be published elsewhere). 
With CGC initial conditions even the QGP phase must exhibit significant 
dissipative effects if one want to reproduce the RHIC data.

In Figs.~\ref{F3} and \ref{F4} we show the measured \cite{data} 
rapidity dependence of $v_2$ for three centrality classes, together
with calculations at three representative impact parameters $b\eq4,\,6.3\,$
and 8.5\,fm corresponding to these centrality selections (i.e. adjusted
to give the correct average number of participants $N_{\mathrm{part}}$
in each case as quoted in Ref.~\cite{data}). Results from BGK initial 
conditions are shown in Fig.~\ref{F3}. The lines show ideal fluid 
dynamical calculations with kinetic decoupling assumed at 
$T_{\rm dec}\eq100$\,MeV (solid blue) and 
$T_{\rm dec}\eq{T}_\mathrm{sw}\eq169$\,MeV
(dashed green). The dashed lines underpredict the data at all 
impact parameters and all but the most forward rapidities, indicating 
the need for generating additional elliptic flow during the hadronic 
stage below $T_c$. The solid lines, on the other hand, strongly 
overpredict the forward rapidity data in semiperipheral and peripheral
collisions, showing that ideal hydrodynamics generates too much additional
elliptic flow during the hadronic stage. The hydro+cascade hybrid model
(red circles) gives a good description of the data over the entire 
rapidity range for all three centralities, with the exception of the 
midrapidity region in the most central collision sample as already
discussed above. The hadronic cascade model provides just the right
amount of dissipation to bring the ideal fluid prediction down to
the measured values, especially in very peripheral collisions and
away from midrapidity.       

The situation is different for the more eccentric CGC initial conditions, 
as shown in Fig.~\ref{F4}. Now $v_2$ is overpredicted at all centralities 
and rapidities if the hadronic phase is described by ideal fluid dynamics,
and the dissipative effects of the hadronic cascade are no longer 
sufficient to reduce $v_2$ for the more peripheral bins enough to 
obtain agreement with the data. In the peripheral sample (25-50\%) the 
excess elliptic flow persists at almost all rapidities and exists even 
if the fireball freezes out directly at hadronization (no hadronic 
evolution at all). Significant dissipation in the early QGP phase is 
needed in this case to bring the theoretical prediction in line with 
experiment.

We conclude that the answer to the question, whether all of the 
observed discrepancies between elliptic flow data and ideal fluid 
dynamical simulations can be blamed on ``late hadronic viscosity''
and fully eliminated by employing a hydro+cascade hybrid model such
as the one studied here, depends on presently unknown details of the
initial state of the matter formed in the heavy-ion collision. 
With BGK initial conditions hadronic dissipation seems to be able
to reduce the elliptic flow enough to bring the theoretical 
predictions in line with the data, leaving little room for 
additional dissipative effects in the early QGP stage. CGC initial 
conditions yield significantly more eccentric sources and produce
larger than observed elliptic flow even if dissipative effects in
the late hadronic stage are taken into account. In this case, the 
early QGP phase must either be significantly softer than parametrized 
by our equation of state or exhibit significant viscosity itself.
With CGC initial conditions, the excess over the $v_2$ data from 
Au+Au collisions at RHIC persists even at midrapidity in all but 
the most central collisions; with such initial conditions, the 
standard claim \cite{QGP3} that at RHIC energies the measured 
elliptic flow at midrapidity exhausts the theoretical upper limit 
predicted by ideal fluid dynamics must be qualified.

We see that a data-based attempt to establish limits on the viscosity of 
the quark-gluon plasma requires, among other things, a better understanding
of the initial conditions of the fireball created in RHIC collisions.
Unfortunately, very few direct probes of the initial conditions are
available. In Ref.~\cite{AG05,AGH05} 3-dimensional jet tomography
was proposed to test the longitudinal structure of BGK and CGC 
initial conditions in noncentral collisions. A specific feature
of CGC initial conditions is a predicted sign flip of the first 
Fourier moment $v_1$ of nuclear modification factor $R_{AA}(p_T,y,\phi)$ 
at high $p_T$ as one moves away from midrapidity \cite{AGH05}. 
Alternatively, one can try to exploit the fact that the large (even 
if not perfect) degree of thermalization observed in heavy-ion 
collisions at RHIC limits the amount of entropy produced during 
the expansion. When taking into account that final state rescatterings 
in the medium produce only very small effects on the shape of the 
rapidity distribution, the finally observed charged hadron rapidity 
distributions therefore severely constrain the initial entropy and
energy density profiles \cite{KLN01}. A better theoretical 
understanding of the initial conditions, especially of the transition 
from the high density to small density regimes, is needed to extract 
the viscosity of quark--gluon plasma at the early stages. A systematic 
study of the charged hadron rapidity distributions for a variety of 
collision centralities, center of mass energies and system sizes is 
needed to assess which description of the initial state yields a 
more consistent and efficient overall description of all available 
data.

\acknowledgments
This work was supported by the U.S. DOE under contracts 
DE-FG02-93ER40764 (T.H.), DE-FG02-01ER41190 (U.H.),
DE-AC02-98CH10886 (D.K.)
and DE-FG02-87ER40331.A008 (R.L.).
Discussions with A.~Adil, M.~Gyulassy, 
and A.J.~Kuhlman are gratefully acknowledged. 

%%%%%%%%%%%%%%%%%%%%%%%%  References %%%%%%%%%%%%%%%%%%%%%%%%%%%%%%%%%%%%%%%%%

%\bibliography{references}

\begin{thebibliography}{99}
\bibitem{fn2}
  The inclusion of event-by-event fluctuations in both the geometric 
  shape and density profile of the initial entropy distribution is 
  numerically expensive, and for our hydro+cascade hybrid model it 
  exceeds the available technical resources. However, their role in 
  affecting the extraction of $v_2$ from theoretical calculations
  and experimental data is presently being intensely discussed 
  \cite{GRZ,Soco,Hama,MS03}. We point out in particular the work
  by the Sao Paulo -- Rio de Janeiro Collaboration \cite{Soco,Hama}
  who include such fluctuations in their ideal fluid dynamical model
  and find that they tend to reduce the hydrodynamically predicted
  elliptic flow, especially at forward rapidities. Whether this might 
  help to reduce the discrepancy between theory and data for our
  hydro+cascade hybrid with CGC initial conditions remains to be 
  studied.
\end{thebibliography}

\end{document}